\documentclass[aps,prl,twocolumn,floatfix,superscriptaddress,showpacs,showkeys]{revtex4-1}
\usepackage{epsf,amsmath,amssymb,verbatim,color,multirow,pifont}
\usepackage{graphicx}

\newcommand{\Smo}{Smoluchowski }
\newcommand{\tp}{{\hat t}}

\newcommand{\SR}{{S_{R}}}

\begin{document}
\setcounter{page}{1}
\title[]{Hybrid Percolation Transition in Cluster Merging Processes: \\ Continuously Varying Exponents}
\author{Y.S.~\surname{Cho}}
\affiliation{CCSS,  CTP and  Department of Physics and Astronomy, Seoul National University, Seoul 08826, Korea}
\author{J.S.~\surname{Lee}}
\affiliation{School of Physics, Korea Institute for Advanced Study, Seoul 02455, Korea}
\author{H.J.~\surname{Herrmann}}
\affiliation{Computational Physics for Engineering Materials, Institute for Building Materials, ETH Z\"urich, 8093 Z\"urich, Switzerland}
\author{B.~\surname{Kahng}}
\email{bkahng@snu.ac.kr}
\affiliation{CCSS,  CTP and Department of Physics and Astronomy, Seoul National University, Seoul 08826, Korea}
%\date[]{Received \today}

\begin{abstract}
Consider growing a network, in which every new connection is made between two disconnected nodes. At least one node is chosen randomly from a subset consisting of $g$ fraction of the entire population in the smallest clusters. Here we show that this simple strategy for improving connection exhibits a phase transition barely studied before, namely a hybrid percolation transition exhibiting the properties of both first-order and second-order phase transitions. The cluster size distribution of finite clusters at a transition point exhibits power-law behavior with a continuously varying exponent $\tau$ in the range $2 < \tau(g) \le 2.5$. This pattern reveals a necessary condition for a hybrid transition in cluster aggregation processes, which is comparable to the power-law behavior of the avalanche size distribution arising in models with link-deleting processes in interdependent networks. 
\end{abstract}

\maketitle
Transport or communication systems grow by adding new connections. Often certain constraints are imposed by society and if these constraint involve global knowledge about connectivity, the transition to a percolating system can become first order, as happens for instance when suppressing the spanning cluster, when imposing a cluster size~\cite{science} or when favoring the most disconnected sites~\cite{gaussian}. Typically this effect is accompanied by the loss of critical scaling making these abrupt transitions less predictable and thus more dangerous. We will show here, that for a specific case, namely a variant of the model introduced in Ref.~\cite{half}, critical fluctuations and power-law distributions can prevail and for the first time identify a hybrid transition in explosive percolation. 

Hybrid phase transitions have been observed recently in many complex network systems~\cite{rev1,rev2}; in these transitions, the order parameter $m(t)$ exhibits behaviors of both first-order and second-order transitions simultaneously as   
\begin{equation}
m(t)=\left\{
\begin{array}{lr}
0 & ~{\rm for}~~  t < t_c, \\
m_0+r(t-t_c)^{\beta} & ~{\rm for}~~ t \ge t_c, 
\end{array}
\right.
\label{order}
\end{equation}
where $m_0$ and $r$ are constants and $\beta$ is the critical exponent of the order parameter, and $t$ is a control parameter. Examples of such behavior include $k$-core percolation~\cite{kcore1,kcore2}, the cascading failure model on interdependent complex networks~\cite{havlin,mendes}, and the Kuramoto synchronization model with a correlation between the natural frequencies and degrees of each node on complex networks ~\cite{moreno, mendes_sync}, etc. For the models in~\cite{kcore1,kcore2,havlin,mendes}, a critical behavior appears as nodes or links are deleted from a percolating cluster above the percolation threshold until reaching a transition point $t_c$. As $t$ is decreased infinitesimally further as $1/N$ beyond $t_c$ in finite systems, the order parameter decreases suddenly to zero and a first-order phase transition occurs. Thus, a hybrid phase transition occurs at $t=t_c$ in the thermodynamic limit. 
%Note that in the cascading failure model in multiplex networks, the criticality at $t_c$ is induced by the presence of a power-law avalanche size distribution of finite clusters separate from an infinite cluster~\cite{havlin} when a node or link is deleted from the giant mutually connected component in multiplex networks. On the other hand, the size distribution of mutually connected components at $t_c$ does not follow a power law.  

Next we recall discontinuous percolation transitions occurring in generalized contagion models~\cite{grassberger_2012,janssen,dodds}. Recent studies~\cite{chung} of a generalized epidemic model \cite{janssen} revealed that the discontinuous percolation transition turns out to be a hybrid percolation 
transition (HPT) represented by (\ref{order}). For this case, a HPT is induced by cluster merging processes. However, the critical behavior arising in a HPT in a cluster merging process has not been yet studied at all, even though one may guess that its nature can differ from that of the HPT in link-deleting processes~\cite{kcore1,kcore2,havlin,mendes}. This Letter aims to understand the nature of an HPT in a cluster merging process  and identify the similarities and differences in the phase transition compared with those of HPTs  in link-deleting processes. Our study is based on a simple stochastic model introduced later, from which we could obtain analytic solutions for diverse properties of the critical behavior. 
%Moreover, the results and methods we obtained and used would not be limited to 
%the model itself but would be applicable to further research on HPTs  in cluster merging processes, such as those in synchronization models \cite{moreno,mendes_sync}, or a generalized epidemic model~\cite{janssen,chung}.

\begin{figure}[t]
\includegraphics[width=.90\linewidth]{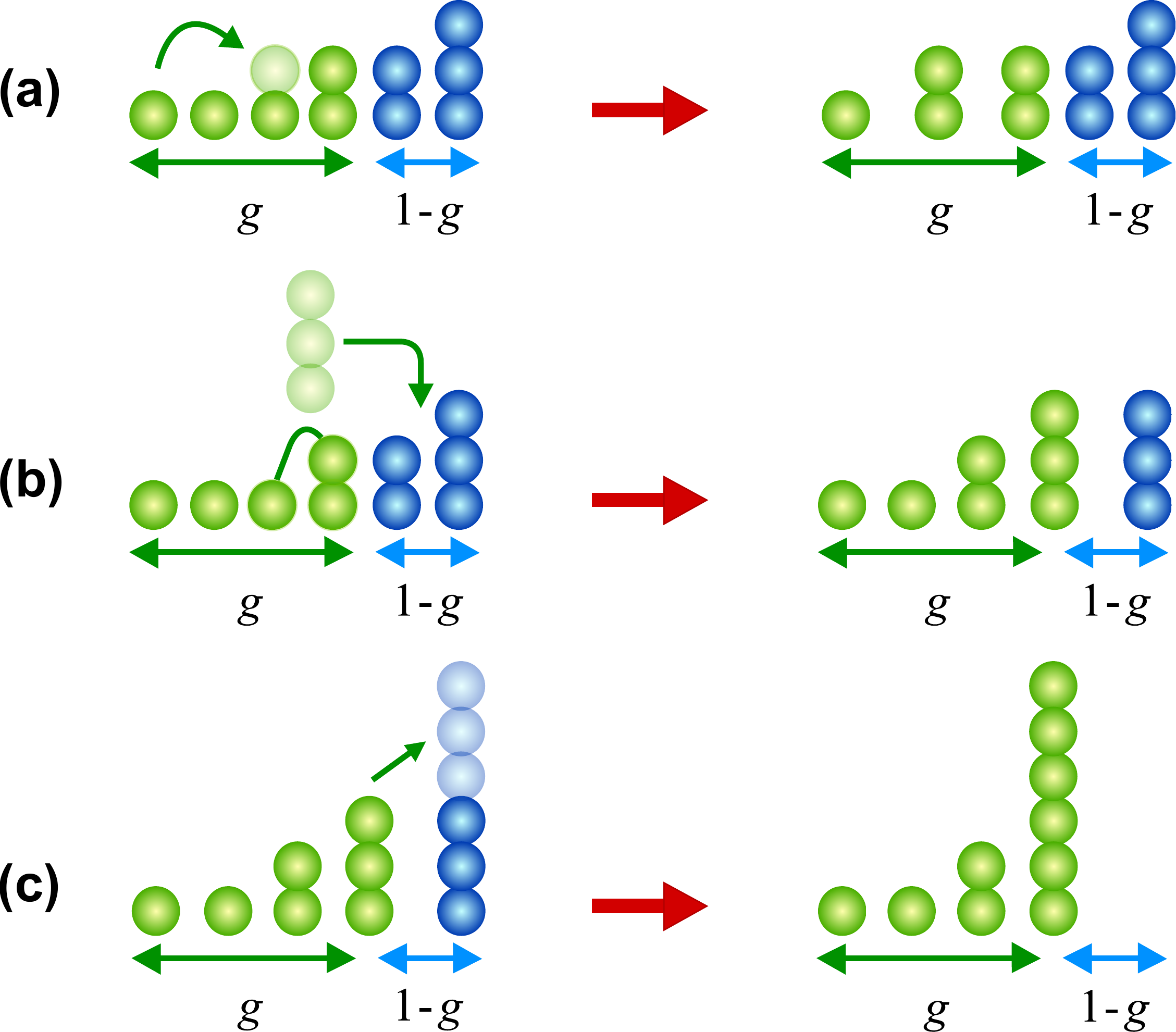}
\caption{ (Color online) Schematic illustration of the r-ER model with $g=0.5$ and $N=10$.
Nodes (represented by balls) in $R(t)$ are green (light gray), whereas those in $R^{(c)}(t)$ are  blue (dark gray). Columns  represent clusters. At each time step, two nodes are selected, one from the green balls and the other from balls of any color. (a) When the two balls are both green, if the sum of their cluster sizes is smaller than or equal to $S_R(t)$, the size of the largest cluster in $R(t)$, then $S_R(t)$ remains the same. 
(b) However, if the sum is larger than $S_R(t)$, then $S_R(t)$ can be increased. 
The cluster of two blue balls no longer belongs to $R^{(c)}(t)$ but moves to $R(t)$. The merged cluster is 
still in $R(t)$ because it contains a ball with ranking $gN$. 
(c) When the size of the largest cluster in the system exceeds $(1-g)N$ for the first time, which is referred to as $t_g$, all nodes are regarded as green balls. After this happens, the dynamics follows that of the ordinary 
ER model.   
}
\label{fig_model}
\end{figure}

 The model we study is defined as follows: We start with a system consisting of $N$ isolated nodes. At each time step, one node is selected uniformly at random from the entire system and the other node is selected from a restricted set consisting of approximately $g$ fraction of the entire nodes, as defined below. The two nodes are connected by an edge unless they are already connected. The time step $t$ is defined as the number of edges added to the system per node, which is a control parameter. The restricted set is defined as follows: First, we rank clusters by ascending order of size at each time step. When more than one cluster of the same size exists, they are sorted randomly. At this stage, a restricted set of clusters $R(t)$ is defined as the subset consisting of a certain number of the smallest clusters (say $k$ clusters), denoted as $R(t) \equiv \{c_1, c_2, \cdots, c_k\}$. Further, $k$ is determined as the value satisfying the inequalities, $N_{k-1}(t) < \lfloor gN \rfloor \le N_k(t)$ for a given model parameter $g\in (0,1]$. $N_k(t) \equiv \sum_{\ell=1}^k s_{\ell}(t)$, where $s_{\ell}(t)$ is the number of nodes in cluster $c_{\ell}$. We note that the number of clusters in $R(t)$ varies with time. For later discussion, we denote the number of nodes in the set $R(t)$ as $N_R(t)\equiv N_k(t)$ and the size of the largest cluster in $R(t)$ as $S_R(t)$. 
This model is called a restricted Erd\H{o}s and R\'enyi (r-ER) model, because when $g=1$, the model is reduced to percolation in the ordinary ER model.  The model is depicted schematically in Fig.\ref{fig_model}. We remark that this r-ER model is a slightly modified version of the original model~\cite{half} in which the number of nodes in the set $R$ is always $\lfloor gN \rfloor$, independent of time. Thus, when $N_{k-1}(t) < \lfloor gN \rfloor < N_k(t)$, some nodes in one of the largest clusters in $R$ belong to 
$R$ and the other nodes of the same cluster belong to $R^{(c)}$. However, in our modified model, all the nodes in that cluster belong to the set $R$. This modification enables us to solve analytically the phase transition for $t > t_c$ without changing any critical properties (see supplementary information (SI)) 

\begin{figure}[h]
\includegraphics[width=.92\linewidth]{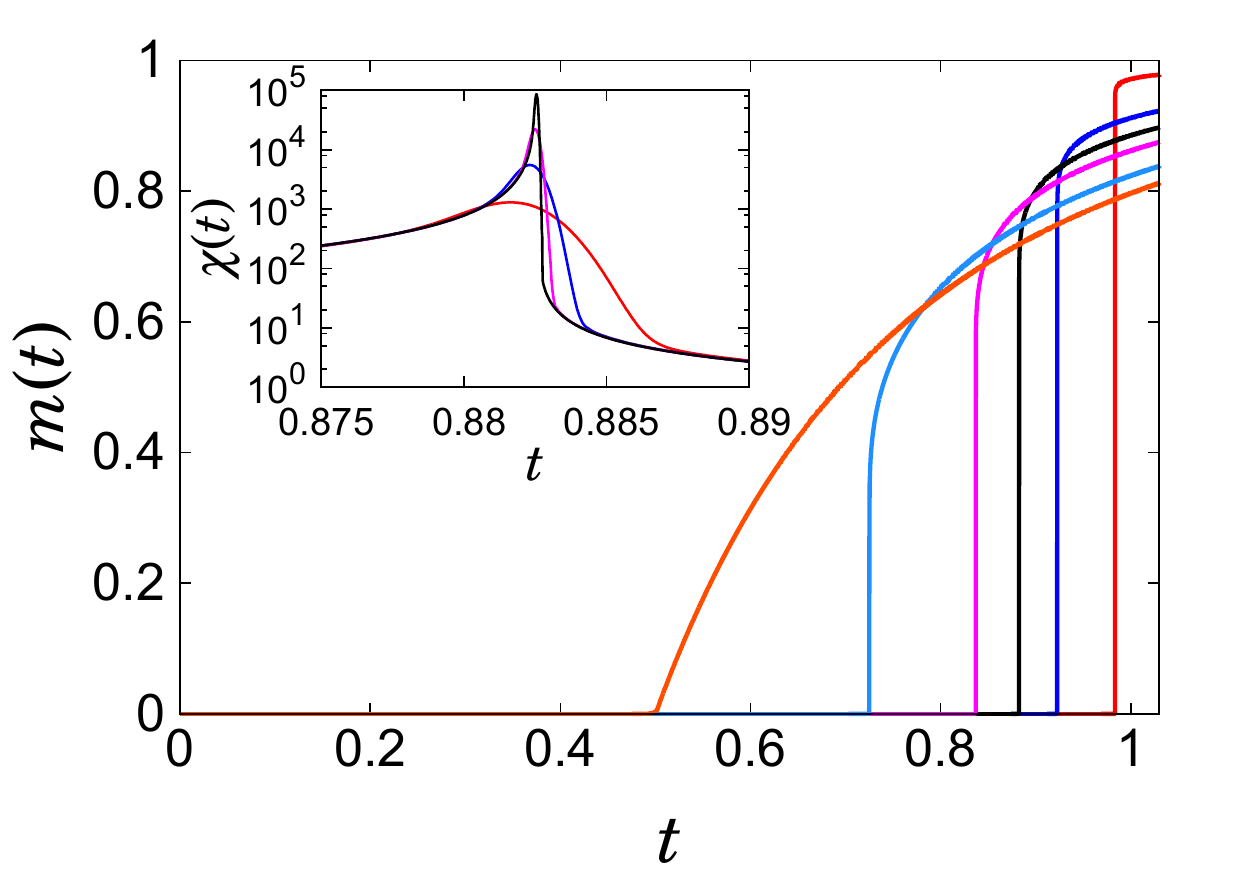}
\caption{(Color online) Plot of $m(t)$ vs $t$ for different model parameter values $g=0.2, 0.4, 0.5, 0.6, 0.8$ and 1.0 from right to left. 
Inset: Plot of the susceptibility $\chi(t)$ vs $t$ for different system sizes $N/10^4=8, 64, 512~ {\rm and}~ 4096$ for a fixed $g=0.5$. 
}
\label{fig_order}
\end{figure}

\begin{figure}[h]
\includegraphics[width=1.0\linewidth]{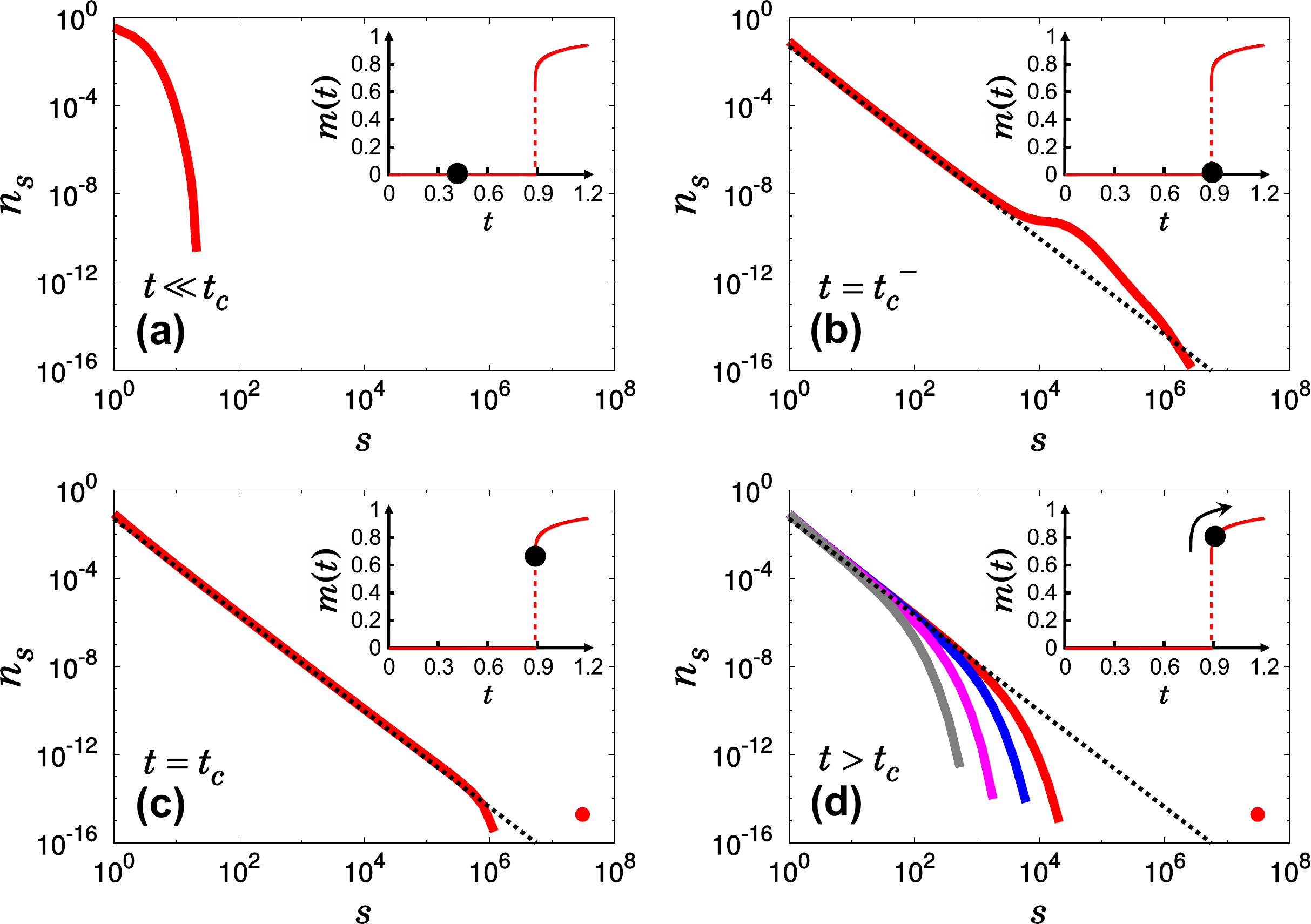}
\caption{(Color online) Plot of $n_s(t)$ vs $s$ at several time steps, (a) $t \ll t_c$, (b) $t=t_c^-$, (c) $t=t_c$, and (d) $t > t_c$. Inset in each panel indicates how the order parameter $m(t)$ behaves at each time.}
\label{fig_clustersize}
\end{figure}

The r-ER model exhibits an HPT at a percolation threshold $t_c$, which is delayed compared with the percolation threshold of the ordinary ER model as shown in Fig.~\ref{fig_order}, This behavior is similar to that obtained in the explosive percolation model \cite{explosive,bfw,science,natphy}. The  order parameter is the fraction of nodes belonging to the giant cluster, denoted as $m(t)$. This order parameter begins to increase abruptly from a certain tipping point $t_c^-$ defined in \cite{tipping} (see also Fig.\ref{fig_clustersize}b) and reaches a finite value $m_0 < 1$ at $t_c$. Thus during the interval $\Delta t \equiv t_c -t_c^-$, the order parameter increases drastically. The interval $\Delta t$ scales as $o(N)/N$~\cite{half}, which reduces zero in the limit $N\to \infty$. Thus, the abrupt transition becomes a discontinuous transition in the thermodynamic limit. This property is also confirmed using finite-size scaling analysis, which is presented in the SI.  
After $t_c$, the order parameter $m(t)$ increases gradually. Here we argue that in general, continuously increasing behavior of $m(t)$ for $t > t_c$ does not guarantee an HPT.  One needs to check whether the order parameter increases continuously following Eq.~(\ref{order}), and other physical quantities follow critical behaviors. 

We examine the cluster size distribution $n_s(t)$ for several time steps as shown in Fig.~\ref{fig_clustersize}, 
where $n_s(t)$ denotes the number of clusters of size $s$ divided by $N$.   
i) When $t \ll t_c$,  $n_s(t)$ decays exponentially with respect to $s$. 
ii) When $t= t_c^-$, $n_s(t)$ exhibits power-law decay in the small-cluster region but contains a bump in the large-cluster region. 
iii) When $t=t_c$, at which the order parameter becomes $m_0$, $n_s(t)$ exhibits power-law decay with the exponent $\tau$ for finite clusters, at an $s$-distance from a giant cluster positioned separately at $s=m_0N$. Note that $\tau$ depends on the model parameter $g$.
iv) When $t > t_c$, the size distribution of finite clusters exhibits crossover behavior: it undergoes power-law decay for $s < s^*$, but exponential decay for $s> s^*$. Thus, $n_s(t) \sim s^{-\tau} e^{-s/s^*},$ where $s^*\sim (t-t_c)^{-1/\sigma}$ according to the conventional scaling theory. 
 
We consider the evolution of the cluster size at an arbitrary time step $t$ using the \Smo equations for the cluster size distribution $n_s(t)$.
\begin{widetext}
\begin{eqnarray}
\frac{dn_s(t)}{dt}&=&\sum_{i,j=1}^{\infty} \frac{{i n_i}{j n_j}}{g}\delta_{i+j,s}-\Big(1+\frac{1}{g}\Big)sn_s  \hskip 7cm \textrm{for}~ s < \SR, 
\label{eq:1st} \\
\frac{dn_s(t)}{dt}&=&\sum_{i,j=1}^{\infty}\frac{{i n_i}{j n_j}}{g}\delta_{i+j,s}-sn_s-\Big(1-\sum_{k=1}^{\SR-1}\frac{in_i}{g}\Big) \hskip 5.5cm \textrm{for}~ s=\SR,  
\label{eq:2nd}\\
\frac{dn_s(t)}{dt}&=&\sum_{j=1}^{\infty}\delta_{i+j,s}jn_j\sum_{i=1}^{\SR-1}\frac{{i n_i}}{g}+
\sum_{j=1}^{\infty}\delta_{\SR+j,s}jn_j\Big(1-\sum_{i=1}^{\SR-1}\frac{{i n_i}}{g}\Big)-sn_s  \hskip 2.6cm \textrm{for}~s > \SR.
\label{rate}
\end{eqnarray}
\end{widetext}
Note that the number of clusters of size $S_R$ can be greater than one. In this case, some of them are randomly chosen to belong to $R(t)$ and the others belong to $R^{(c)}$ 
to satisfy the inequalities $N_{k-1}(t) < \lfloor gN \rfloor \le  N_k(t)$ presented previously. Thus, we need to consider the case $s=\SR$ separately. 
In the above equations, we used the approximation that the total number of nodes in the set $R$ is $N_R \approx gN$ for $t< t_g$, where $t_g$ is the time step at which the size of the largest cluster in the system exceeds $(1-g)N$ for the first time. Thus, $g$ is taken as unity for $t \ge t_g$. Further, $t_g$ is located between $t_c^-$ and $t_c$. The derivations of each term for each case of $s$ are explained in the SI.  Moreover, performing direct numerical integration of the \Smo equations, we successfully reproduce the behaviors of $m(t)$ in Fig.\ref{fig_order} and $n_s$ in Fig.\ref{fig_clustersize}, which are shown in the SI. 

We focus on the rate equation for $n_1(t)$ as follows: 
At early initial time steps, monomers are abundant and can be in both $R$ and $R^{(c)}$. 
However, as time passes, their number decreases and at a certain time step denoted as $t_1$, all the monomers are in the set $R$. The rate equation for monomers is written separately as 
$\frac{dn_1}{dt}= -n_1-1$ for $t < t_1$ using Eq.(\ref{eq:2nd}), and $\frac{dn_1}{dt}= -\left(1+\frac{1}{g}\right)n_1(t)$ for $t > t_1$ using Eq.({\ref{eq:1st}). Then 
\begin{equation}
n_1(t) = \left\{\begin{array}{lll}
2e^{-t}-1 &  {\rm if}~~~t < t_1, \\
g(\frac{g+1}{2})^{-(1+\frac{1}{g})}e^{-(1+\frac{1}{g})t} & {\rm if}~~~t > t_1,
\end{array}\right.\label{canstep}
\end{equation}
and $t_1$ is obtained as $t_1={\textrm {ln}}(\frac{2}{g+1})$. 

Next, we use the fact that $n_s(t_c)$ follows a power law, i.e., $n_s(t_c)=a_0s^{-\tau}$ for $1\le s \le S_0$, where $S_0$ is the size of the largest cluster among the finite clusters and $a_0$ can be determined in terms of $g$ and $t_c$ from Eq.~(\ref{canstep}). 
We can also use the relation $\sum_{s=1}^{\infty}n_s(t_c)=1-t_c$, because up to $t_c$, links are 
likely to be added between clusters, which is checked numerically in the SI. We also confirm that $d\sum_{s=1}^{\infty} n_s(t)/dt=-1$ from the \Smo equations (\ref{eq:1st}-\ref{rate}). 
Moreover, the relation $\sum_{s=1}^{S_0}sn_s(t_c) \approx \sum_{s=1}^{\infty} sn_s(t_c)=1-m_0$ can be 
used in the limit $N\to \infty$. Taking account of those facts, we obtain the self-consistency equation for the exponent $\tau$ as   
\begin{equation}
\frac{\zeta(\tau)}{\zeta(\tau-1)} =\frac{1}{1-m_0}\left\{1-{\textrm {ln}}\left(\left[\frac{g\zeta(\tau-1)}{1-m_0}\right]^{\frac{g}{1+g}}\left(\frac{2}{g+1}\right)\right)\right\}.
\label{originoftau}
\end{equation}
To obtain $\tau$, one needs $m_0$ values for given $g$ values. We use the numerically 
estimated values of $m_0$ in Table \ref{tauestimate}.
The obtained values denoted as $\tau$ range between $2 < \tau(g) \le 2.5$, as listed in Table \ref{tauestimate}. We note that as $g$ decreases, $\tau$ becomes smaller, and the jump in the order parameter becomes larger. This result implies that the growth of large clusters is more strongly suppressed at smaller $g$. 

To obtain the exponent $\sigma$ analytically, we take into account two facts: First, the dynamics in regime iv) is of the ER type: Because $m_0>1-g$, the restricted set covers the entire system. Next we take $t_c$ as an $ad$ $hoc$ time origin, and the cluster size distribution at $t_c$ is used as an initial condition of the ER dynamics. Under this transformation, the solution for the evolution of the ER model remains the same~\cite{can}.
Then using the solution for the cluster size distribution of the ER model~\cite{ziff}, we obtain that $sn_s(\tp)=s^{1-\tau}f(s\tp^{1/\sigma})$, where $f(x)$ is a scaling function and 
$\tp\equiv t-t_c$ for $t> t_c$. Using the property that $f(x)$ is an analytic function~\cite{mendes_initial}, 
we obtain $\sigma=1$ independent of $g$. The detailed derivation is presented in the SI. 

The order parameter $m(t)$ increases continuously with time 
when $t>t_c$. To study the criticality for $t> t_c$, we again take the transition point $t_c$ as an {\it ad hoc} time origin. Next, we use the formalism for the ordinary ER model with an arbitrary initial cluster size distribution as presented in \cite{can, ziff, mendes_initial}. 
For instance, the order parameter can be obtained via the self-consistency equation, $m(\tp)=H(2tm(\tp))$, where $H(\mu) \equiv 1-\sum_s sn_s(0)e^{-\mu s}$ \cite{can,ziff}. Using $n_s(\tp=0)=a_0s^{-\tau}$ with $a_0=(1-m_0)/\zeta(\tau-1)$, we obtain that $m(\tp=t-t_c)=m_0+r (t-t_c)^{\tau-2}$,
where $r=-\frac{\Gamma(2-\tau)}{\zeta(\tau-1)}(1-m_0)(2m_0)^{\tau-2}$.
This gives $\beta=\tau-2$. The detailed derivation of the exponent $\beta$ is presented in the SI. 
Note that we used $m(\tp)=m_0+r\tp^{\beta}$ instead of $m(\tp) \sim {\hat t}^{\beta}$ to obtain the above result. If we had used $m(\tp)\sim \tp^{\beta}$, relevant to the case $m_0=0$, we would have obtained $\beta=(\tau-2)/(3-\tau)$. For the ER model, $\tau_{\rm ER}=5/2$, and $\beta_{\rm ER}=1$. Accordingly, we say that the discontinuity of the order parameter $m_0 \ne 0$ at $t_c$ 
changes the criticality for $t> t_c$.

The susceptibility $\chi^+ \equiv \sum_{s=1}^{\prime} s^2 n_s(t)$ for $t> t_c$, where the prime indicates the exclusion of an infinite-size cluster, can be obtained using $\chi^+={H^{\prime}(2\tp m(\tp))}/[{1-2\tp H^{\prime}(2\tp m(\tp))}]$ with $\tp=t-t_c$, where the prime in $H(\mu)$ indicates the derivative with respect to its argument $\mu$. When $2< \tau < 3$, $H^{\prime}(\mu)$ has a non-integer singularity of $\mu^{\tau-3}$. Plugging $\mu=2\tp m_0$ into the formula for $\chi^+(t)$, we obtain that $\chi^+(t) \sim (t-t_c)^{-\gamma}$ with $\gamma=3-\tau$. Note that the denominator is finite at $\tp =0$. 

Obtaining the analytical result of the susceptibility $\chi^-(t) \equiv \sum_{s=1}^{\infty}s^2 n_s(t)$ for $t< t_c$ is intriguing. Using the \Smo equation for the r-ER model, we obtain the following relation ${\partial \chi^-(t)}/{\partial t}=2\chi^-(t)\chi_R(t),$ where $\chi_R\equiv (1/g)\sum_{s=1}^{S_R}s^2 n_s(t)$ and we used the approximation $\sum_{s=1}^{S_R} sn_s \approx g$ which is valid near $t_c^-$. If we assume that $\chi^-(t)$ diverges algebraically as $t \to t_c$,  i.e., $\chi^-\sim (t_c-t)^{-\gamma^{\prime}}$, which is confirmed numerically, then we could obtain  
$\chi_R=(\gamma^{\prime}/2) (t_c-t)^{-1}$. However, the exponent $\gamma^{\prime}$ was not analytically determined. 
In fact, $\chi_R$ is another form of the susceptibility defined via the correlation function~\cite{mendes_susc} 
as $\chi_c\equiv (1/N)\sum_{i,j=1}^N C(i,j)$, where $C(i,j)$ is the probability that two nodes $i$ and $j$ belong to the same finite cluster. 
Thus, $C(i,j)$ can be obtained as the probability that two nodes $i$ and $j$ are selected from the same cluster. Then, 
$\chi_c=N \sum_{\alpha \in R}({s_{\alpha}}/{N})({s_{\alpha}}/{gN})$, where $\alpha$ is the index of cluster. 
This formula reduces to $\chi_R$ using $N_R/N\approx \sum_{s=1}^{S_R} sn_s\approx g$ near $t_c^-$.  

A critical behavior at $t_c$ appears in the form of the power-law-type size distribution of finite clusters, which is necessary to generate an HPT, with exponent $\tau$ ranging between $2 < \tau \le 2.5$. This pattern is analogous to the power-law behavior of the avalanche size distribution in the cascading failure model~\cite{havlin}. 
%However, it is not clear yet how this critical behavior of the avalanche size distribution is related to that of the order parameter in the cascading failure model. 
%Therefore, understanding of the nature of HPTs in cluster merging processes is much easier than that of the HPT in link-deleting processes, and the obtained result for the former will be helpful for understanding more precisely the critical behavior of the latter. 

The r-ER model is a simple model exhibiting an HPT in a cluster merging process.  As the classical ER model has served as a basic model for understanding the evolution of social networks, the r-ER model should be similarly useful but under a certain constraint that the least connected members are preferred to connect to others, for instance, as in the merging of the lowest financial companies under government control in financial crisis. Moreover, the theoretical framework obtained from the r-ER model can be used for understanding other HPTs  in cluster merging processes,  for instance, in generalized epidemic models~\cite{janssen,chung} and synchronization models~\cite{moreno,mendes_sync}. Our preliminary results suggest that indeed HPTs in cluster merging processes could be found from those models. 

We have shown here, that in a model, in which the most disconnected agents are preferentially connected, spanning occurs abruptly, while astonishingly critical fluctuations and power-law distributions prevail after the transition within the critical region. These post-transition critical mergers of rather big clusters into a rather meager spanning one represent a rather transparent geometrical interpretation of the recently discovered hybrid phase transitions.

\begin{table*}
\caption{For a given model parameter $g$, percolation threshold ($t_c$), discontinuity of the order parameter at $t_c$ ($m_0$),  the exponent of the cluster size distribution at $t_c$ obtained from Eq.~(\ref{originoftau}) using the mean value $m_0$ ($\tau^*$), numerically obtained ($\tau$), the exponent ($1/\sigma$), the exponent ($\beta$), the exponent ($\gamma^{\prime}$) for $t\to t_c^-$, and ($\gamma$) for $t\to t_c
$ . All simulation results are obtained from several different system sizes up to $N=4 \times 10^7$. The estimated 
values of the critical exponents and error bars are confirmed to be independent of the system sizes. The details of how to 
measure them are presented in the SI.  
%Numerical values are in reasonable agreement with the theoretical predictions, $\beta=\tau-2$, $\gamma=3-\tau$ and $\sigma=1$ when $g$ is not close to unity. However, when $g\to 1$, $\tau^*$ and $\tau$ change drastically to the ER value $\tau_{\rm ER}=2.5$, and all the other exponents also change drastically to the ER values $\sigma_{\rm ER}=1/2$, $\beta_{\rm ER}=1$, and $\gamma^{\prime}_{\rm ER}=\gamma_{\rm ER}=1$.
}
\begin{center}
\begin{tabular}{ccccccccc}
\hline\hline ~~$g$~~~&~~~~~~~~~~$t_c$~~~~~~~~~~ &~~~~~~~$m_0$~~~~~~~&~~~~~$\tau^*$~~~~~&~~~~~~$\tau$~~~~~~&~~~~~~~$1/\sigma$~~~~~~~&~~~~~~~$\beta$~~~~~~~&~~~~~~~$\gamma^{\prime}$~~~~~~~&~~~~~~~$\gamma$~~~~~~~
\\
\hline 
$0.1$ & $0.9984\pm0.0002$ & $0.98\pm0.05$ & $2.06$ & $2.03\pm 0.04$ & $1.00\pm 0.05$ & $0.04\pm 0.05$ & $1.00\pm 0.02$ & $0.97 \pm 0.05$  \\
$0.2$ & $0.9836\pm0.0002$ &$0.90\pm0.05$ & $2.13$ & $2.08\pm 0.04$ & $1.01\pm 0.05$ & $0.09\pm 0.05$ & $1.03\pm 0.02$ & $0.91 \pm 0.05$  \\
$0.3$ & $0.9570\pm0.0002$ &$0.81\pm0.05$ & $2.18$ & $2.12\pm 0.04$ & $1.01\pm 0.05$ & $0.13\pm 0.05$ & $1.06\pm 0.03$ & $0.88 \pm 0.05$  \\
$0.4$ & $0.9227\pm0.0002$ &$0.72\pm0.05$ & $2.23$ & $2.16\pm 0.04$ & $1.02\pm 0.05$ & $0.17\pm 0.04$ & $1.08\pm 0.03$ & $0.85 \pm 0.05$  \\
$0.5$ & $0.8826\pm0.0002$ &$0.62\pm0.05$ & $2.26$ & $2.18\pm 0.04$ & $1.04\pm 0.05$ & $0.21\pm 0.05$ & $1.10\pm 0.03$ & $0.83 \pm 0.05$  \\
$0.6$ & $0.8370\pm0.0002$ &$0.52\pm0.06$ & $2.29$ & $2.20\pm 0.04$ & $1.05\pm 0.05$ & $0.23\pm 0.04$ & $1.11\pm 0.04$ & $0.82 \pm 0.05$  \\
$0.7$ & $0.7852\pm0.0003$ &$0.41\pm0.05$ & $2.32$ & $2.22\pm 0.04$ & $1.08\pm 0.05$ & $0.28\pm 0.05$ & $1.13\pm 0.05$ & $0.81 \pm 0.05$  \\
$0.8$ & $0.7250\pm0.0003$ &$0.29\pm0.05$ & $2.35$ & $2.25\pm 0.04$ & $1.12\pm 0.05$ & $0.32\pm 0.05$ & $1.15\pm 0.05$ & $0.81 \pm 0.05$  \\
$0.9$ & $0.650\pm0.002$ &$0.15\pm0.05$ & $2.38$ & $2.28\pm 0.04$ & $1.20\pm 0.10$& $0.39\pm 0.08$  & $1.15\pm 0.05$ & $0.80 \pm 0.05$  \\
\hline
\end{tabular}
\end{center}
\label{tauestimate}
\end{table*}

This work was supported by NRF grants (No. 2010-0015066 and No. 2014R1A3A2069005), the exchange program between the Korean NRF and ERC, ERC Advanced Grant No. FP7-319968-FlowCCS and the Global Frontier Program (YSC).


\begin{thebibliography}{99}
\bibitem{tipping} $t_c^-$ can be taken as the intercept with the $t$-axis of the tangential line of $m(t)$ at the inflection point of $m(t)$. $t_c^-$ differs from $t_c$ in finite systems, but it reduces to $t_c$ in the thermodynamic limit.
\end{thebibliography}
\end{document}